\documentclass[pdflatex,sn-mathphys-num]{sn-jnl}%

\usepackage{graphicx}%
\usepackage{multirow}%
\usepackage{amsmath,amssymb,amsfonts}%
\usepackage{amsthm}%
\usepackage{mathrsfs}%
\usepackage[title]{appendix}%
\usepackage{xcolor}%
\usepackage{textcomp}%
\usepackage{manyfoot}%
\usepackage{booktabs}%
\usepackage{algorithm}%
\usepackage{algorithmicx}%
\usepackage{algpseudocode}%
\usepackage{listings}%
\usepackage{parskip}
\usepackage{physics}
\usepackage[small,hang,bf]{caption}    %
\usepackage{subcaption}
\usepackage{braket} %
\usepackage{cleveref} %
\usepackage{ulem} %
\def\bajo#1#2{\smash{\mathop{#1}\limits_{#2}}}  %

\theoremstyle{thmstyleone}%

\theoremstyle{thmstyletwo}%

\theoremstyle{thmstylethree}%

\begin{document}

\title[Article Title]{Effect of isotropic errors on the complexity of Grover's algorithm}

\author*[1,2]{\fnm{Anurag} \sur{Saha Roy}}\email{anurag@qruise.com}

\author[2]{\fnm{Jesús} \sur{Lacalle}}\email{jesus.glopezdelacalle@upm.es}

\affil*[1]{\orgname{Qruise GmbH}, \orgaddress{\street{Am H\"olzersbach 7}, \city{Saarbr\"ucken}, \postcode{66113}, \country{Germany}}}

\affil[2]{\orgdiv{ETS de Ingeniería de Sistemas Informáticos}, \orgname{Universidad Politécnica de Madrid}, \orgaddress{\street{C/ Alan Turing s/n}, \city{Madrid}, \postcode{28031}, \country{Spain}}}

\abstract{Isotropic errors have been shown to be immune to conventional error correction techniques. While general theoretical frameworks have been proposed to model such errors, there have been no studies so far analysing their concrete impact on practical use-cases. Here we explore the effect of isotropic errors on the complexity of Grover's search algorithm through numerical simulations, with an analysis of the impact on the algorithm's performance and success probability. The results provide insights into the robustness of Grover's algorithm against isotropic errors, highlighting potential challenges for implementations on noisy quantum hardware. All results presented here are obtained through numerical simulations using the open-source python library \texttt{isotropic} developed as part of this work. The source code, numerical simulations and documentation for the library are available online at \url{https://www.github.com/lazyoracle/isotropic-error-analysis}.}


\maketitle

\section{Introduction}\label{sec:introduction}

The transition of quantum computing from theoretical promise to practical implementation faces significant challenges, with quantum errors emerging as perhaps the most fundamental barrier to realizing the full potential of quantum computation. All contemporary quantum hardware platforms suffer from significant error rates that typically range from 0.01\% to 1\% per gate operation. While these error rates may appear small in isolation, the cumulative effect over the hundreds or thousands of operations required for meaningful quantum algorithms becomes substantial. Quantum error correction represents the most comprehensive approach to addressing quantum noise, employing redundant encoding of logical qubits across multiple physical qubits to detect and correct errors. Pioneering schemes such as the surface code~\cite{fowler2012surface} and color code~\cite{landahl2011fault} have demonstrated theoretical feasibility for fault-tolerant quantum computation. More recently, work from both industry and academia~\cite{google2025quantum,xu2024constant,besedin2026lattice,bluvstein2024logical,Krinner2022,reichardt2024demonstration} has demonstrated that error correction indeed works in practice, successfully suppressing logical error rates to allow the development of large scale devices. The effectiveness of these techniques depends critically on the underlying assumptions regarding the structure and characteristics of the quantum errors they are designed to address. However, recent theoretical work has revealed the existence of error types that fundamentally challenge these assumptions. Among these, isotropic errors represent a particularly intriguing class of quantum noise that exhibits unique mathematical properties and resistance to conventional error correction approaches.

Isotropic errors, as discussed in~\cite{Oliveira_2017,Lacalle_2019,Lacalle_2021,lacalle2026fidelity}, are characterized by their rotational symmetry around quantum states in the high-dimensional Hilbert space representation. Unlike conventional error models that typically affect specific components of quantum states in predictable ways, isotropic errors maintain equal probability of occurrence in all directions around the original quantum state. Recent work~\cite{Lacalle_2021} has demonstrated that quantum codes cannot effectively correct isotropic errors, regardless of the specific code construction or encoding strategy employed. This surprising result stems from the geometric properties of isotropic errors in the projective Hilbert space, where the symmetry of the error distribution prevents the establishment of distinguishable error syndromes that are essential for error correction protocols.

The inability of quantum error correction codes to address isotropic errors highlights a potential vulnerability in our current understanding of quantum error correction and points to the need for alternative approaches to quantum error analysis. While much research~\cite{Scott2005,Preskill2013,Gottesman2014,Cross2009,Hill2013,DuclosCianci2014,Hocker2016a,Hocker2016b,Aliferis2006,Wang2010,Aggarwal2010,Criger2016,Ozen2012,Li2013,Chen2019,LaGuardia2012,Boulant2002,Evans2012,Dias2014,Naghipour2015,Greenbaum2018,Bravyi2018,Piltz2014,Buterakos2018} has focused on characterizing quantum algorithms and quantum error correction thresholds under well-studied error models such as depolarizing noise, amplitude damping, miscalibrated controls or crosstalk noise, the behavior in the presence of isotropic errors remains largely unexplored. This work addresses this knowledge gap by conducting a comprehensive numerical analysis of the impact of isotropic errors on quantum algorithm performance, specifically on the Grover's algorithm for unstructured search. Grover's algorithm serves as an ideal testbed due to its fundamental importance in quantum computing and its well-understood theoretical framework. The algorithm's geometric structure, based on amplitude amplification through repeated rotations, provides a clear framework for understanding how isotropic errors affect the complexity of quantum algorithms.

Our main contribution is presented in~\cref{sec:results}, where we numerically analyse the performance of Grover's algorithm under varying magnitudes of isotropic errors. We demonstrate that even small isotropic errors can significantly degrade the success probability of the algorithm, and our results reveal that while increasing the number of iterations can partially compensate for the effects of isotropic errors, there are fundamental limits to this approach due to the geometric nature of the Grover's algorithm.~\Cref{sec:method} recaps the mathematical representation of isotropic errors, and presents the simulation methodology we use for our numerical analysis, and the software tools~\cite{isotropic2025} developed for this purpose. Finally,~\cref{sec:discussion} synthesizes the findings and discusses their broader implications for quantum computing, including potential avenues for future research in quantum error analysis and algorithm design.

\section{Method}\label{sec:method}

\subsection{Mathematical description of isotropic errors}\label{ssec:mathematical-description-of-isotropic-errors}
Let $\Phi$ be an $n$-qubit quantum state given by the following expression:
\begin{equation}\label{Eq:QubitFormula}
    \Phi = \sum_{x=0}^{2^n - 1} \left( a_{2x} + ia_{2x + 1}\right) \ket{x} \text{such that} \sum_{x=0}^{2^{n+1}-1} a_x^{2} = 1.
\end{equation}

This $n$-qubit state could be represented on a unit sphere of dimension $d = 2^{n+1}-1$ in the $\mathbb{R}^{d+1}$. This unit sphere can then be represented in spherical coordinates as explained below:

\begin{equation}\label{Eq:Hypersphere}
    S_d = \left\{(a_0, a_1, a_2, \dots a_d) \mid a_0^{2} + a_1^{2} + a_2^{2} + \cdots + a_d^{2} = 1\right\}
\end{equation}

Since density matrices do not distinguish between different types of quantum computing errors~\cite{nielsen2002}, we describe quantum computing errors as random variables $X$ with density function defined on $S^d$. As shown in~\cite{Lacalle_2019}, we can obtain the corresponding density matrix of $X$, $\rho(X)$ using the formula:
\begin{equation}\label{Eq:DensityMatrix}
\rho(X)=\int_{S^d}f(x)|\Phi\rangle\langle\Phi|dx\quad\text{where}\quad\int_{S^d}f(x)dx=1.
\end{equation}

We now consider the state $\Psi$ which is the final state caused by an error on $\Phi$. For a given random variable $X$, we can define its variance as the mean of the quadratic deviation from the mean value $\mu$ of $X$, i.e., $V(X)=E[\|X-\mu\|^2]$. When talking about quantum computing errors, this $E(X)$ is the original $n-$qubit $\Phi$ state before an error has occurred. For pure quantum states represented by~\cref{Eq:QubitFormula},~\cite{Lacalle_2019} shows that the variance of $X$ will be
\begin{equation}\label{Eq:Variance}
    V(X)=E[\|\Psi-\Phi\|^2]=E[2-2x_0]=2-2\int_{S^d}x_0f(x)dx.
\end{equation}

We can now define a quantum variance which relates this to the fidelity measure~\cite{nielsen2002} commonly used in quantum computing. Quantum states are equivalent under multiplication by a phase, so the quantum variance is defined as:
\begin{equation}\label{Eq:QuantumVariance}
    V_q(X)=E[\bajo{\text{min}}{\phi}(\|\Psi-e^{i\phi}\Phi\|^2)]=2-2E\left[\sqrt{x_0^2+x_1^2}\right].
\end{equation}

It has been shown~\cite{Oliveira_2017,Lacalle_2019,Lacalle_2021,lacalle2026fidelity} that the quantum variance and fidelity can be used interchangeably for independent errors and when the quantum variance tends to $0$, fidelity tends to $1$ and vice-versa. Thus, we now have a way to relate the random variable with the state fidelity, as caused by an error generating process.

\begin{figure}[ht!]
    \centering
    \includegraphics[width=\textwidth]{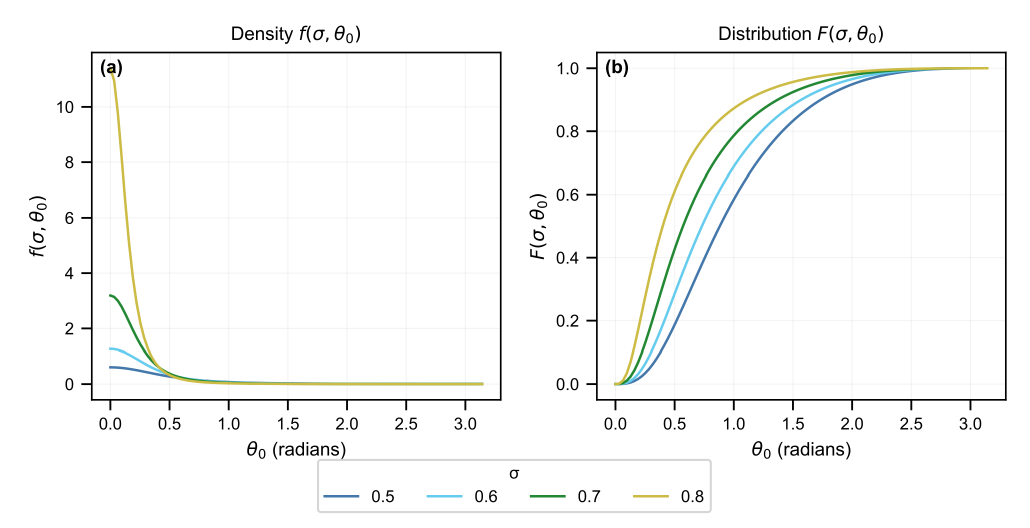}
    \caption{Value of the Density function $f(\sigma,\theta_0)$ (left) and Distribution function $F(\sigma,\theta_0)$ (right) for $d=2$ for different values of $\sigma$.}\label{fig:Distr}
\end{figure}

An isotropic error is one where the probability density function of the error state $\Psi$ depends only on its distance from a fixed point $\Phi$ on the sphere. In other words, the density function is invariant under rotations around the point $\Phi$. This means that the probability of an error occurring at a certain distance from $\Phi$ is the same in all directions. When modelling errors on a qubit, we consider the following unitary matrix $E$ that parameterises an arbitrary element of the group $SU(2)$ by three angles $(\theta_0, \theta_1, \theta_2)$, such that $\theta_0, \theta_1 \in [0,\pi]$ and $\theta_2 \in [0,2\pi]$:

\begin{equation}
    \begin{pmatrix}
    \cos\theta_0 + i\sin\theta_0\cos\theta_1 & \sin\theta_1(-\sin\theta_0\cos\theta_2+i\sin\theta_0\sin\theta_2) \\
    \sin\theta_1(\sin\theta_0\cos\theta_2+i\sin\theta_0\sin\theta_2) & \cos\theta_0 - i\sin\theta_0\cos\theta_1
    \end{pmatrix}
\end{equation}

Thus, $\Psi = E\Phi$ and for a 1-qubit state, we can show that $\norm{\Psi - \Phi}^2 = 2 - 2 \cos\theta_0$, i.e., the distance between $\Phi$ and $\Psi$ depends only on $\theta_0$ and not on $\theta_1$ or $\theta_2$ or the state $\Phi$. Here, we deal with a special class of isotropic errors called the normal error distribution which has the following density function:

\begin{equation}\label{Eq:NormalDistr}
    f(\sigma,\theta_0)=\frac{(2d-2)!!}{(2\pi)^d}\frac{(1-\sigma^2)}{(1+\sigma^2-2\sigma\cos(\theta_0))^d}%
\end{equation}

where the parameter $\sigma$ belongs to the interval $[0,1)$.%

As we can see in~\cref{Eq:NormalDistr}, the density function depends only on $\theta_0$ and not on $\theta_1$ or $\theta_2$, which is a requirement for isotropic errors. The parameter $\sigma$ controls the spread of the distribution around the point $\Phi$. As $\sigma$ tends to $1$ the probability gets concentrated around the point $\Phi$, essentially eliminating the error. And when $\sigma$ approaches $0$ the distribution tends to be uniform, i.e., it causes all the states to be equally likely. \Cref{fig:Distr} shows how the distribution changes depending on the parameter $\sigma$.

\subsection{Properties of isotropic errors}\label{ssec:properties-of-isotropic-errors}
A key property of isotropic errors that makes their simulation relatively straightforward is that they commute with quantum gates. In effect, if an isotropic error $E$ occurs and we then apply a quantum gate (unitary operator) $U$, the result is the same as if we first applied the gate $U$ and then an error $E'$ occurs with the same probability distribution as $E$, i.e., $U\odot E = E' \odot U$. This property allows us to apply all the quantum errors that occur during a computation to the final state, just before performing the measurement of the qubits that allows us to obtain the algorithm's result.

Furthermore, isotropic errors possess an additional property that also simplifies the analysis. The sum of isotropic errors is also an isotropic error. If the errors we sum follow normal distributions (see \cref{Eq:NormalDistr}), with parameters $\sigma_1, \sigma_2, \dots, \sigma_j$, the resulting error also follows a normal distribution with a parameter:

\begin{equation}\label{Eq:sigma-series}
    \sigma_f = \sigma_1 \times \sigma_2 \times \cdots \times \sigma_j
\end{equation}

If all the intermediate errors follow a normal distribution with the same parameter $\sigma$ and the algorithm length (number of gates) is $j$, then the total computational error can be simulated by a single error at the end of the computation (just before the qubit measurement) with a normal distribution and parameter:

\begin{equation}\label{Eq:sigma-scaled}
    \sigma_f = \sigma^j
\end{equation}

\subsection{Numerical simulation of isotropic errors}\label{ssec:numerical-simulation-of-isotropic-errors}

\begin{algorithm}[ht!]
\caption{Algorithm to generate and apply isotropic error}\label{alg:isotropic-error}
\begin{algorithmic}[1]
    \State\textbf{Input:} statevector $\Phi$ and isotropic error variance $\sigma$
    \State\textbf{Output:} statevector $\Psi$ with isotropic error applied
    \Function{IsotropicError}{$\Phi, \sigma$}
        \State$d \gets 2^{n+1}-1$
        \State$\{\mathbf{v}^1, \mathbf{v}^2, \dots, \mathbf{v}^d\} \gets$ orthonormal basis of $\Pi$
        \State$\{\theta_1, \theta_2, \dots, \theta_{d-1}\} \gets$ angles with density function $f_j(\theta_j)$ as in~\cref{Eq:DensityFunctionTheta}
        \State$\mathbf{e}_2 \gets$ vector in $S_{d-1}$ with coefficients as in~\cref{Eq:e2}
        \State$\theta_0 \gets$ angle with density function $f(\theta_0)$ as in~\cref{Eq:NormalDistr}
        \State$\Psi \gets \Phi \cos(\theta_0) + e_2 \sin(\theta_0)$
        \State\Return$\Psi$
    \EndFunction %
\end{algorithmic}
\end{algorithm}

Given a quantum state $\Phi$, we can generate an isotropically perturbed state $\Psi$ by following the steps outlined in algorithm~\ref{alg:isotropic-error}. The algorithm takes as input the original state $\Phi$ and the isotropic error variance $\sigma$, and outputs the perturbed state $\Psi$. The steps of the algorithm are explained in detail below.

We want to create an orthonormal basis of the hyperplane $\Pi$ that is tangential to the hypersphere $S_{d-1}$ with center at $\Phi$. The orthonormal basis consists of a set of $d$ vectors in $\mathbb{R}^{d+1}$, i.e., each vector has $d+1$ coordinates, is of unit magnitude, and is orthogonal to both $\Phi$ and all other vectors in the basis. Since this depends on the state $\Phi$, the orthonormal basis is obtained in terms of the coordinates of $\Phi$, i.e., the spherical representation presented in~\cref{Eq:Hypersphere}. We follow the steps below to build this orthonormal basis incrementally:

\begin{enumerate}
    \item First vector with only two non-zero coordinates $\mathbf{v}^1 = (x_0, x_1)$ such that the homogeneous linear system $\mathbf{a_0 x_0 + a_1 x_1 = 0}$ is fulfilled. Solving the coefficient matrix by the Gauss method gives:
    \[
    (a_0 \ a_1) \implies
    \mathbf{v}^1 =
    \begin{cases}
    \dfrac{(a_1, -a_0)}{\sqrt{a_0^2 + a_1^2}} & \text{if } a_0 \ne 0 \text{ or } a_1 \ne 0 \\
    (1, 0) & \text{if } a_0 = a_1 = 0
    \end{cases}
    \]
    \item Second vector with only three non-zero coordinates $\mathbf{v}^2 = (x_0, x_1, x_2)$ such that the following homogeneous linear system is fulfilled and $\mathbf{x_0^2 + x_1^2 + x_2^2 = 1}$:
    \[
    \begin{cases}
    v_0^1 x_0 + v_1^1 x_1 & = 0 \\
    a_0 x_0 + a_1 x_1 + a_2 x_2 & = 0
    \end{cases}
    \implies
    \begin{pmatrix}
    v_0^1 & v_1^1 & 0 \\
    a_0 & a_1 & a_2
    \end{pmatrix}
    \]
    \item And, in general, $\mathbf{k}$-th vector with only $\mathbf{k+1}$ non-zero coordinates $\mathbf{v}^k = (x_0, x_1, \ldots, x_k)$ such that the following homogeneous linear system is fulfilled and $x_0^2 + x_1^2 + \cdots + x_k^2 = 1$:
    \begin{equation}\label{Eq:OrthonormalBasisMatrix}
    \begin{pmatrix}
    v_0^1 & v_1^1 & 0 & 0 & \cdots & 0 & 0 \\
    v_0^2 & v_1^2 & v_2^2 & 0 & \cdots & 0 & 0 \\
    \vdots & \vdots & \vdots & \vdots & \ddots & \vdots & \vdots \\
    v_0^{k-1} & v_1^{k-1} & v_2^{k-1} & v_3^{k-1} & \cdots & v_{k-1}^{k-1} & 0 \\
    a_0 & a_1 & a_2 & a_3 & \cdots & a_{k-1} & a_k
    \end{pmatrix}
    \end{equation}

\end{enumerate}

The matrix has rank $k-1$ (because the vectors $\mathbf{v}^1, \dots, \mathbf{v}^{k-1}$ are orthonormal) or $k$ (if $a_k \ne 0$) and $k+1$ unknowns. Therefore it is an indeterminate compatible system and a unit vector $\mathbf{v}^k$ can be obtained as a solution which together with $\mathbf{v}^1, \dots, \mathbf{v}^{k-1}$ forms an orthonormal system. At the end of this first step, we have obtained a set of $d$ orthonormal vectors $\mathbf{v}^1, \mathbf{v}^2, \dots, \mathbf{v}^d$ each having $d+1$ coordinates.

Now that we have an orthonormal basis, we need the coefficients that will be pre-multiplied with each basis vector to obtain the vector $\mathbf{e}_2$ with a uniform distribution on the sphere $S_{d-1}$. This requires $d-1$ angles $\theta_1, \dots, \theta_{d-1}$ such that $\theta_j \in [0, \pi]$ for all $1 \le j \le d-2$ and $\theta_{d-1} \in [0, 2\pi]$. Then we obtain the vector $\mathbf{e}_2$ whose coefficients are as follows in coordinates with respect to the orthonormal basis of $\Pi$ obtained in the previous step:
\begin{equation}\label{Eq:e2}
\mathbf{e}_2 =
\begin{bmatrix}
\cos(\theta_1) \\
\sin(\theta_1) \cos(\theta_2) \\
\sin(\theta_1) \sin(\theta_2) \cos(\theta_3) \\
\vdots \\
\sin(\theta_1) \sin(\theta_2) \dots \sin(\theta_{d-2}) \cos(\theta_{d-1}) \\
\sin(\theta_1) \sin(\theta_2) \dots \sin(\theta_{d-2}) \sin(\theta_{d-1})
\end{bmatrix}
\end{equation}

We thus need to calculate the $d-1$ angles in order to obtain the $d$ elements of the vector $\mathbf{e}_2$. The distribution of $\theta_{d-1}$ is uniform in the interval $[0, 2\pi]$ and for the rest of the angles $\theta_j$, for all $1 \le j \le d-2$, the density function in the interval $[0, \pi]$ is:
\begin{equation}\label{Eq:DensityFunctionTheta}
f_j(\theta_j) = C_j \sin^{d-j-1}(\theta_j) \quad \text{where } C_j =
\begin{cases}
\dfrac{(d-j-1)!!}{2(d-j-2)!!} & \text{if } j \text{ is odd} \\
\dfrac{(d-j-1)!!}{\pi(d-j-2)!!} & \text{if } j \text{ is even}
\end{cases}
\end{equation}

The corresponding distribution functions $F_j(\theta_j) = \int_0^{\theta_j} C_j \sin^{d-j-1}(\theta) \, d\theta$, for $1 \le j \le d-2$, are as follows.

\begin{enumerate}
    \item If $j$ is odd:
    \begin{equation}
    \begin{split}
    F_j(\theta_j) &= \frac{C_j (d-j-2)!!}{(d-j-1)!!} \theta_j - C_j \cos(\theta_j) \\
    &\sum_{k=0}^{(d-j-2)/2} \frac{(d-j-2)(d-j-4)\dots(2k+2)}{(d-j-1)(d-j-3)\dots(2k+1)} \sin^{2k}(\theta_j)
    \end{split}
    \end{equation}

    \item If $j$ is even:
    \begin{equation}
    \begin{split}
    F_j(\theta_j) &= \frac{C_j (d-j-2)!!}{(d-j-1)!!} \theta_j - C_j \cos(\theta_j) \\
    &\sum_{k=1}^{(d-j-1)/2} \frac{(d-j-2)(d-j-4)\dots(2k+1)}{(d-j-1)(d-j-3)\dots(2k)} \sin^{2k-1}(\theta_j)
    \end{split}
    \end{equation}
\end{enumerate}

\noindent
\textbf{Note:} If in $(d-j-2)(d-j-4)\dots(2k+2)$ or $(d-j-2)(d-j-4)\dots(2k+1)$ the first factor is less than the last one, the product is $1$.

The angle $\theta_{d-1}$ has a uniform distribution and is generated without problems, but for the others, between $\theta_1$ and $\theta_{d-2}$, we use the bisection algorithm~\cite{BurdenFaires2015} together with the normal distribution functions described in~\cref{Eq:NormalDistr} for each of the angles. Calculating the distribution function involves evaluation of double factorials, which can quickly get intractable. To recap, the double factorial of a non-negative integer $n$ is defined as:
\begin{equation}
n!! =
\begin{cases}
n \cdot (n-2) \cdot (n-4) \cdots 2 & \text{if } n \text{ is even} \\
n \cdot (n-2) \cdot (n-4) \cdots 1 & \text{if } n \text{ is odd} \\
1 & \text{if } n = -1 \text{ or } n = 0
\end{cases}
\end{equation}

In the accompanying software library \texttt{isotropic}~\cite{isotropic2025}, we provide an optimized implementation that calculates the double factorial ratio without evaluating the individual double factorials directly, which allows us to efficiently compute the distribution functions for large values of $d$.

The vector $\mathbf{e}_2$ in $S_{d-1}$ is then generated by multiplying the coefficients with the orthonormal bases. We must also generate the angle $\theta_0$ with a normal distribution, whose density function is as follows:
\begin{equation}
f(\theta_0) = \frac{(d-1)!!}{(2\pi)^{(d+1)/2}} \frac{1 - \sigma^2}{(1 + \sigma^2 - 2\sigma \cos(\theta_0))^{(d+1)/2}}%
\end{equation}
The corresponding distribution function, after performing the integration with respect to the angles $\theta_{d-1}, \ldots, \theta_1$, is:
\begin{equation}
F(\theta_0) = \frac{(d-1)!!(1-\sigma^2)}{\pi(d-2)!!} \int_0^{\theta_0} \frac{\sin^{d-1}(\theta)}{(1 + \sigma^2 - 2\sigma \cos(\theta))^{(d+1)/2}} \, d\theta%
\end{equation}

The calculation of the function $F(\theta_0)$ is quite complicated, so to simulate the angle $\theta_0$, we use the Simpson's method~\cite{Simpson1737} to approximate the integral. In summary, to calculate the angle $\theta_0$ with the given density function, we proceed as follows:
\begin{enumerate}
    \item We generate a random number $x$ with a uniform distribution in the interval $[0, 1]$.
    \item Apply the bisection algorithm to solve $x = F(\theta_0)$, performing the necessary evaluations of $F$ using Simpson's method.
\end{enumerate}

Finally, we obtain the perturbed state $\Psi$ by using $\mathbf{e}_2$ (consisting of the orthonormal basis and corresponding coefficients) and the angle $\theta_0$ using the expression in Algorithm~\ref{alg:isotropic-error}.

\subsection{Grover's algorithm}\label{ssec:grover-s-algorithm}
Given an unsorted database of $N$ elements, the unstructured search problem is to find a specific item that satisfies a particular condition, typically encoded as a boolean function $f(x)$ that returns 1 for the desired item and 0 otherwise. In classical computing, this problem requires examining each element sequentially until the target is found, resulting in an average complexity of $O(N/2)$ and worst-case complexity of $O(N)$. This linear scaling becomes prohibitive for large datasets, motivating the search for more efficient approaches. Grover's algorithm~\cite{grover1996} can locate the marked item in approximately $O(\sqrt{N})$ operations, representing a significant theoretical advantage.

\begin{algorithm}[ht!]
\caption{Grover's Algorithm for unstructured search}\label{alg:grover}
\begin{algorithmic}[1]
    \State\textbf{Input:} Oracle function $U_f$ that marks target items, database size $N = 2^n$, number of marked items $M$, number of iterations $k$. %
    \State\textbf{Output:} Index of a marked item with high probability. %
    \Function{Grover}{$U_f, N, M, k$}
        \State Initialize $n = \log_2 N$ qubits to state $\ket{0}^{\otimes n}$ %
        \State Apply Hadamard gates to all qubits: $\ket{\psi} = \frac{1}{\sqrt{N}} \sum_{x=0}^{N-1} \ket{x}$ %
        \For{$i = 1$ \textbf{to} $k$}
            \State Apply oracle $U_f$ to mark target states %
            \State Apply diffusion operator $U_s = 2\ket{\psi}\bra{\psi} - I$ %
        \EndFor %
        \State Measure the qubits to obtain an index $x$ %
        \State\Return$x$
    \EndFunction %
\end{algorithmic}
\end{algorithm}

The problem formulation assumes the existence of an oracle function $U_f$ that can recognize the marked item. This oracle acts as a black box that flips the sign of the amplitude corresponding to the marked state while leaving all other amplitudes unchanged. Amplitude amplification forms the theoretical foundation underlying the algorithm. The core principle involves the systematic rotation of the quantum state vector in a two-dimensional subspace spanned by the marked and unmarked states. Each iteration increases the amplitude of the marked states while decreasing the amplitude of the unmarked states, effectively concentrating probability mass on the desired outcomes. The number of iterations must be chosen carefully to maximize success probability. Too few iterations result in insufficient amplification, while too many iterations cause the system to rotate past the optimal point, decreasing success probability. Each Grover iteration typically requires $O(n)$ gates, and with $O(\sqrt{N})$ iterations needed, the total circuit depth scales as $O(n\sqrt{N}) = O(\sqrt{N} \log N)$. See Algorithm~\ref{alg:grover} for a high level outline of the constituent steps.

\subsection{Simulating isotropic errors in Grover's algorithm}\label{ssec:simulating-isotropic-errors-in-grover-s-algorithm}

For all the simulations, we consider the case of a single search item. We define success probability as the probability of measuring the marked state at the end of the circuit. Without any loss of generality, the marked state is always chosen to be $\ket{0\dots11}$ in our case. Once we have the circuit, we extract the pre-measurement statevector which gives us the error-free success probability $p(n)$. For a given problem size $n$, we have a gate depth $N(n)$ of the circuit, where $N(n)$ is defined as the gate depth of a single Grover iteration times the optimal number of iterations for that problem size. $N(n)$ is used to scale the per quantum gate error parameter $\sigma$ to $\sigma_f = \sigma^{N(n)}$. We then add this isotropic error to this statevector using the method described in~\cref{ssec:numerical-simulation-of-isotropic-errors} and again look at the probability of measuring the $\ket{0\dots11}$ state. This gives us the success probability in the presence of isotropic errors, $p_e(n)$. We repeat this process for different values of error level $\sigma$, of problem size $n$ and of number of Grover iterations to see how the success probability changes with respect to these parameters.

To study the practical cost of introducing this error on the algorithm's complexity, we study the number of repetitions $k(n)$ of the algorithm with error necessary to achieve the same success probability as in the error-free case. This can be obtained as follows:

\begin{equation}\label{Eq:k-vs-n}
    {(1 - p_\text{e}(n))}^{k(n)} = 1 - p(n) \implies k(n) = \frac{\log(1 - p(n))}{\log(1 - p_\text{e}(n))}
\end{equation}

where the $k(n)$ represents the increase in the algorithm's complexity due to the presence of isotropic errors, as a function of the problem size.

\subsection{The \texttt{isotropic} software library}\label{ssec:the-isotropic-software-library}
As part of this work, we also release the open source software library \texttt{isotropic}~\cite{isotropic2025} that provides efficient implementations of the algorithms described in~\cref{ssec:numerical-simulation-of-isotropic-errors} to generate isotropic errors and apply them to quantum states. The library is implemented in Python and includes functions for generating isotropic errors with specified variance, applying these errors to quantum states, and analyzing the resulting state fidelity. As of now, only Grover's algorithm is implemented (using Qiskit~\cite{qiskit_javadi_2024}) and analysed here. However, the library accepts an arbitrary quantum state and returns the perturbed quantum state, both as a standard complex array. This makes it easily integrable with other existing quantum computing frameworks such as Cirq~\cite{cirq_developers_2025} and PennyLane~\cite{pennylane_bergholm_2018} that have their own routines for statevector simulation of different quantum algorithms. The source code, documentation, and examples for using the library are available online at \url{https://www.github.com/lazyoracle/isotropic-error-analysis}.

\section{Results}\label{sec:results}

\begin{figure}
    \centering
    \includegraphics[width=0.5\textwidth]{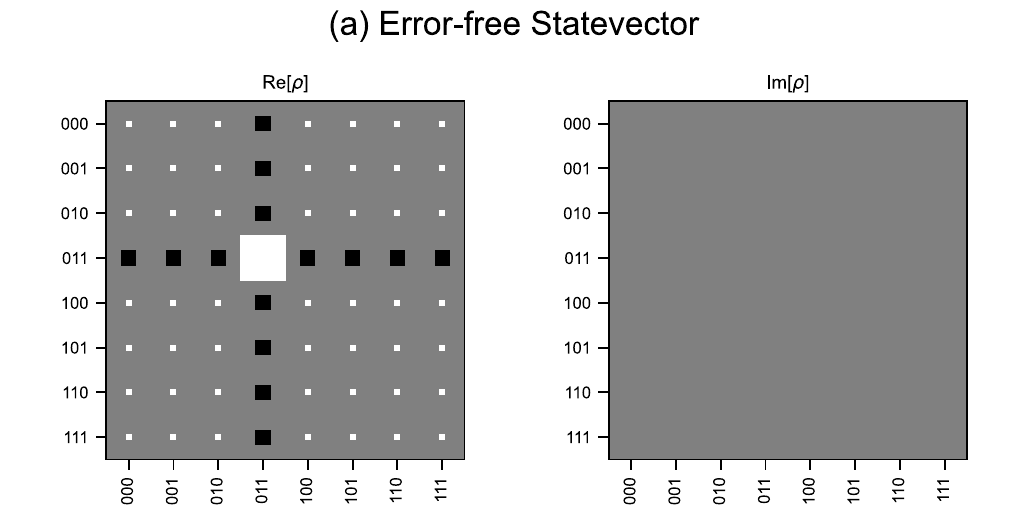}\hfill
    \includegraphics[width=0.5\textwidth]{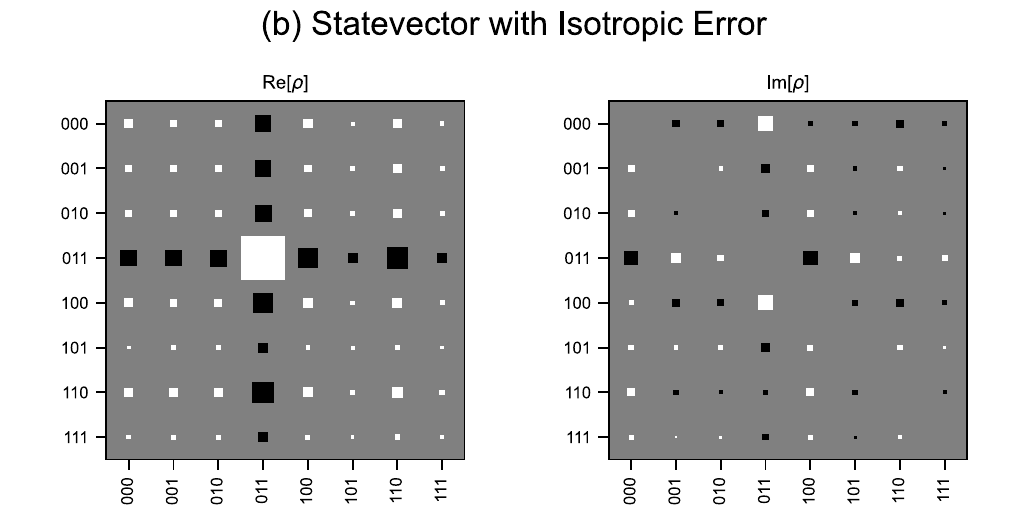}
    \caption{Hinton diagrams of the final statevector before (left) and after (right) applying an isotropic error with $\sigma = 0.999$ for a 3-qubit Grover circuit. We see that while the marked state $\ket{011}$ is still visible after adding the error, the population is now spread across many other states, which reduces the overall success probability.}\label{fig:hinton-diagrams}
\end{figure}

In order to get an understanding of how the isotropic error affects the final statevector of the Grover's algorithm, we first look at~\cref{fig:hinton-diagrams} showing the Hinton diagrams of the final statevector before and after applying an isotropic error with $\sigma = 0.999$ for a 3-qubit Grover circuit. The Hinton diagram is a graphical representation of the statevector where the size of each square represents the magnitude of the corresponding amplitude and the color represents its sign. We see that while the marked state $\ket{011}$ is still visible after adding the error, the population is now spread across many other states, which reduces the overall success probability. For context, $\sigma=0.999$ translates to an approximate gate infidelity of $0.1\%$. This illustrates how even a small isotropic error can significantly degrade the performance of Grover's algorithm by reducing the amplitude of the marked state and increasing the amplitude of unmarked states, thus requiring more repetitions of the algorithm to recover the error free success probability.

\begin{figure}[ht!]
    \centering
    \includegraphics[width=\textwidth]{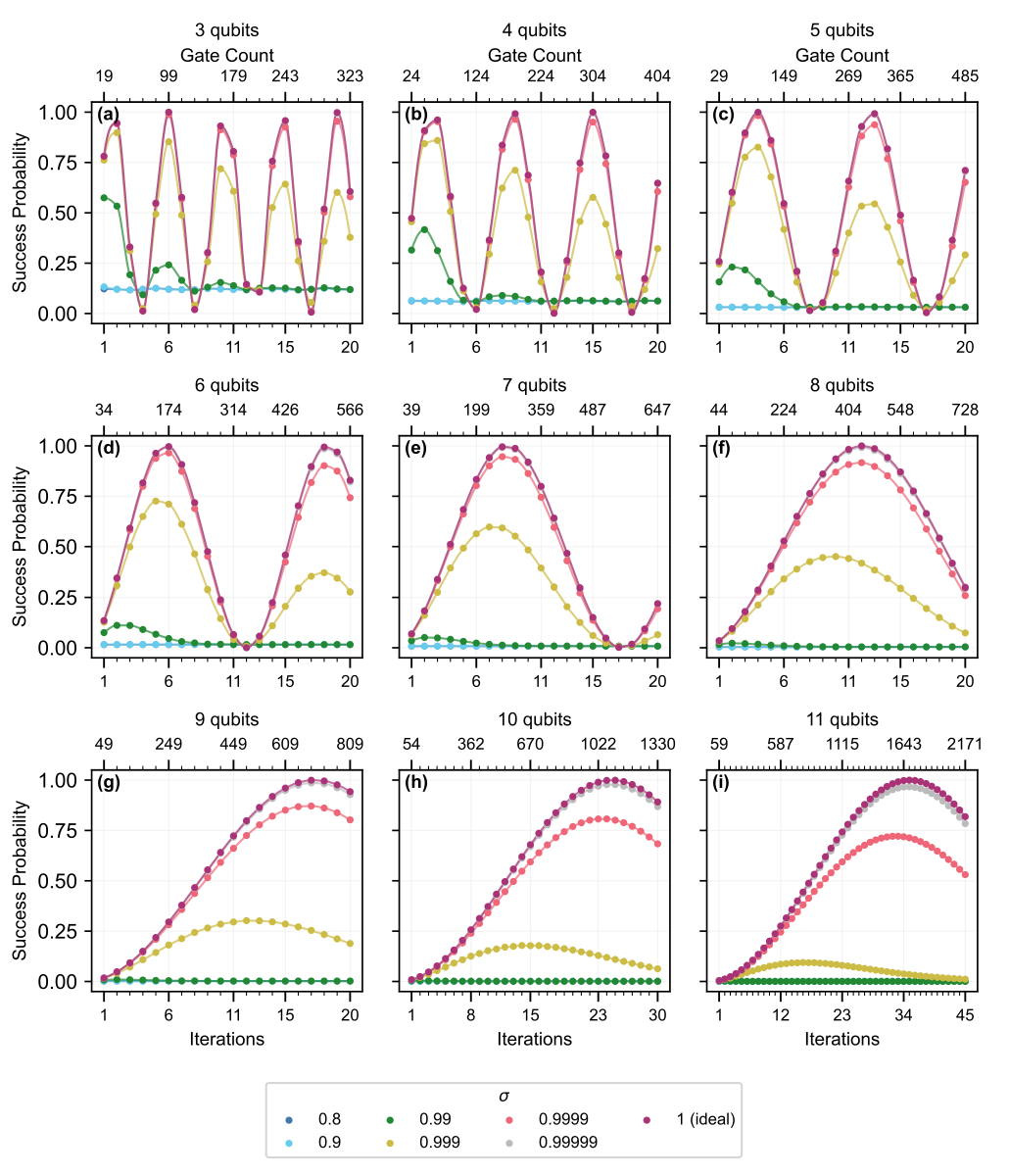}
    \caption{Success probability versus number of Grover iterations for different system sizes and isotropic error magnitudes. Each curve on each subplot represents a different value of $\sigma$, showing how the success probability evolves with the number of iterations for a given error strength.}\label{fig:success-prob-vs-iterations}
\end{figure}

We repeat the process described in \cref{ssec:simulating-isotropic-errors-in-grover-s-algorithm} for system sizes ranging from $3$ to $11$ qubits (corresponding to a problem size of $2^3 = 8$ to $2^{11} = 2048$) and for $\sigma = \{0.8, 0.9, 0.99, 0.999, 0.9999, 0.99999\}$. For each problem size, we look at a broad range of Grover iterations and the results from this analysis are shown in~\cref{fig:success-prob-vs-iterations}. For low error levels, i.e., $\sigma = \{0.99999, 0.9999\}$, the typical Grover oscillation is well-preserved across all qubit counts. Peak success probability remains high, with only a small reduction relative to $\sigma = 1.0$. For moderate error levels $(\sigma = 0.999)$, the peak is visibly attenuated for larger qubit counts ($8 - 11$ qubits), where the higher gate counts amplify the per-gate error. The oscillation shape is still identifiable. For high error levels ($\sigma = \{0.99, 0.9, 0.8\}$), the success probability collapses to near the random-guessing baseline, for all qubit counts starting from 7, indicating that coherent amplification has been completely destroyed by the isotropic errors. At smaller qubit counts ($3 - 5$), partial structure is still visible even at $\sigma = 0.9$. The widening gap between the ideal and noisy curves at larger qubit counts reflects the impact of circuit depth (gate count) which manifests in a lower optimal number of iterations (compared to the ideal case) for certain error levels. In~\cref{fig:success-prob-vs-iterations}, this is clearly visible for $\sigma = 0.999$ for qubit counts $6 - 11$.

\begin{figure}[ht!]
    \centering
    \includegraphics[width=\textwidth]{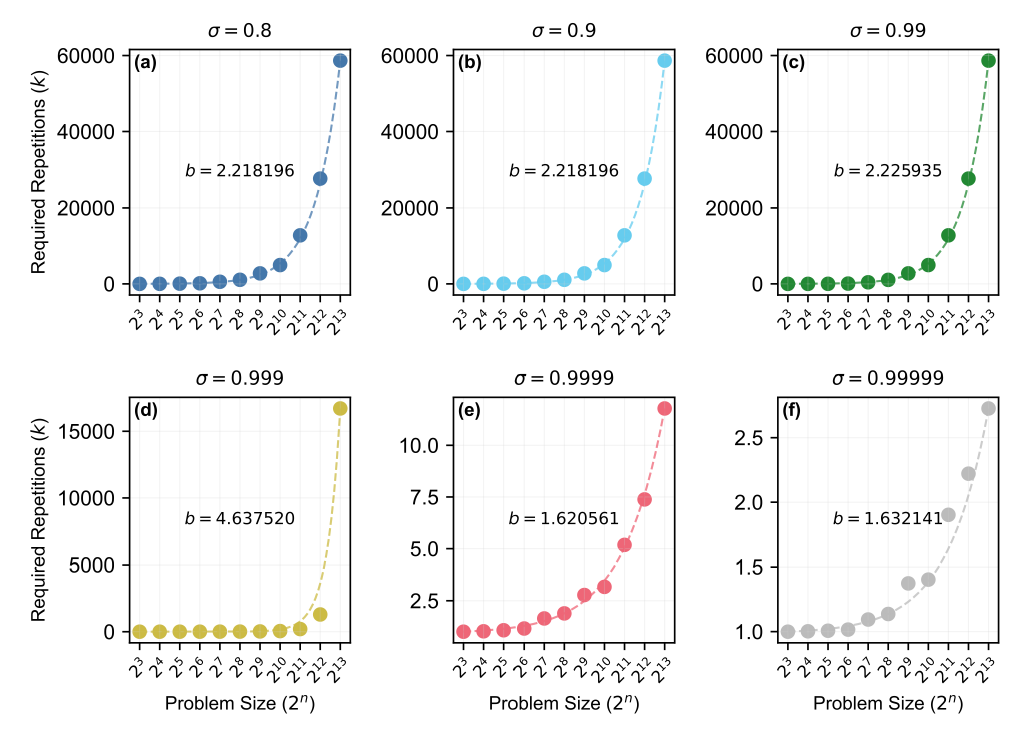}
    \caption{Additional number of repetitions required to achieve the error-free success probability as a function of problem size for different isotropic error magnitudes.}\label{fig:k-vs-problem-size}
\end{figure}

The practical implication of this degradation in success probability is understood through~\cref{Eq:k-vs-n} which outlines a method to calculate the additional number of repetitions that will be required to achieve the error-free success probability in the presence of isotropic errors. This analysis is presented in~\cref{fig:k-vs-problem-size}. Each panel overlays a dashed exponential-linear fit $k(n) = a \cdot b^n + c$, where the constants $a$, $b$, $c$ are determined by exact 3-point interpolation at $n = 3$, $n = 8$, and $n = 13$. The growth rate $b$ characterises how rapidly the repetition overhead increases per additional qubit and is annotated in each panel. The few outlier points we see in this plot arise from the fact that the ideal error-free success probability is not the same for all qubit counts. Correspondingly, the repetition value depends not just on the error level but also shows small deviations based on qubit count. This becomes apparent at high sigma levels because the deviations are no longer relatively small compared to the value of k.

For $\sigma \geq 0.9999$, $k$ remains below $\sim10$ across all problem sizes. Near-ideal performance is recoverable with a small constant repetition overhead, and the overhead grows slowly. When we move on to $\sigma = 0.999$, $k$ starts small but then quickly grows beyond $10$ qubits. For $\sigma \leq 0.99$ $k$ reaches $\mathcal{O}(10^4)$ for $13$ qubits. At these noise levels the success probability at the theoretically optimal iteration count collapses to the random-guessing baseline $(1/N)$.

\begin{figure}
    \centering
    \includegraphics[width=0.8\textwidth]{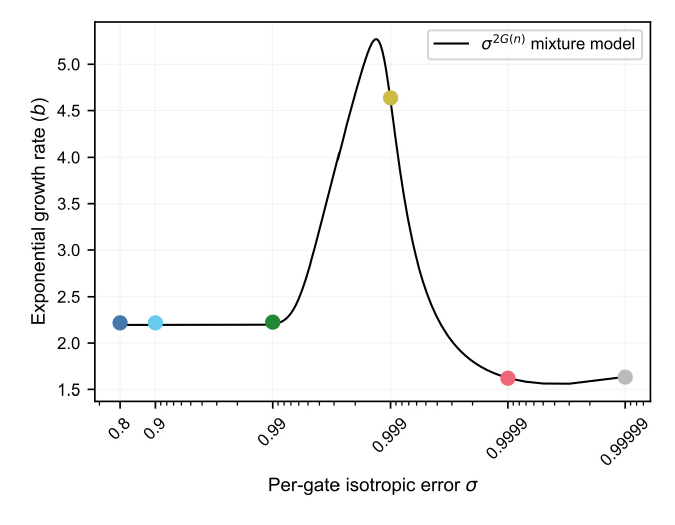}
    \caption{Growth rate $b$ as a function of isotropic error magnitude $\sigma$ overlaid with predicted curve from theoretical mixture model based on total gate count.}\label{fig:b-vs-sigma}
\end{figure}

The exponential growth rate $b$ extracted from each fit characterises how rapidly the repetition overhead $k$ increases per additional qubit.~\Cref{fig:b-vs-sigma} shows $b$ as a function of per-gate fidelity $\sigma$.
We then consider a zero-free-parameter theoretical model motivated by the effect of the total gate count. It treats the circuit as a mixture of coherent and fully decoherent operations, weighted by the cumulative error level $\sigma^{2G(n)}$:

\begin{equation}
    p_e(n) = \sigma^{2G(n)}\, p(n) + \bigl(1 - \sigma^{2G(n)}\bigr) / N
\end{equation}

where $G(n)$ is the total gate count at the optimal number of Grover iterations. The factor $2$ in the weight $\sigma^{2G(n)}$ comes from the squaring necessary to calculate the population probabilities from the amplitudes of the statevector. When $\sigma^{2G(n)} \approx 1$ the circuit is coherent and $p_e \approx p$; when $\sigma^{2G(n)} \approx 0$ the circuit is fully decoherent and $p_e \approx 1/N$ (random guessing). Applying the same exponential-linear fit to the model-predicted $k(n)$ yields the theory curve overlay for $b(\sigma)$ as shown in \Cref{fig:b-vs-sigma}. Because $G(n)$ is read directly from the simulation data, the overlaid curve is a genuine prediction and not a fit; yet it tracks the scatter across the entire $\sigma$ range, reproducing both the flat high-fidelity tail and the rapid low-fidelity rise. Recall that the parameter $b$ is the multiplicative factor by which the required repetitions $k$ grow with each added qubit. Equivalently $k \sim b^{\,n} = N^{\log_2 b}$ (with $N = 2^n$), so $\log_2 b$ is the effective power-law exponent of the overhead in problem size $N$.

\section{Discussion}\label{sec:discussion}
The results presented in this work reveal several important insights into the behavior of Grover's algorithm under isotropic errors, with broader implications for error analysis in quantum algorithms. Our simulations demonstrate that isotropic errors impose a significant penalty on Grover's algorithm, even at relatively modest magnitudes.~\Cref{fig:success-prob-vs-iterations} makes it clear that choosing the optimal number of iterations of the Grover operator depends on the error magnitude, as larger circuit depths accumulate more errors and negate the amplitude amplification. The need for additional repetitions of the algorithm can, to a certain degree, compensate for error-induced amplitude degradation, but the apparent exponential growth of this number of repetitions, illustrated in~\cref{fig:k-vs-problem-size} negates the quadratic speed-up of Grover's algorithm.

Some limitations of this study warrant mention. First, our simulations assume that the isotropic errors are at the same level for different gates in a circuit. This is usually not the case in current generation QPUs where single qubit gates usually have lower error rates compared to entangling operations. Second, we focused exclusively on the single-marked-item case in Grover's algorithm. The multi-solution case and variants such as amplitude estimation may exhibit different sensitivity to isotropic errors and merit separate investigation. Third, our study examined system sizes up to 13 qubits; extending these simulations to larger systems would help establish scaling laws. Finally, while our \texttt{isotropic} library provides efficient implementations of the error generation routine, the computational cost of these simulations remains substantial for large systems. Developing efficient numerical techniques or analytical approximations for specific parameter regimes would facilitate more extensive studies.

\section{Conclusion}\label{sec:conclusion}

We hope the release of our open source library \texttt{isotropic} will enable researchers to easily incorporate isotropic error models into their quantum algorithm simulations and analyses, providing a valuable tool for understanding the impact of such noise on quantum computation. Future work could expand the simulations to larger systems or explore the application of isotropic error models to a wider range of quantum algorithms, including those for optimization, machine learning, and quantum chemistry. Additionally, investigating the interplay between isotropic errors and other types of noise could yield insights into error mitigation and correction strategies. Finally, experimental validation of our findings on real quantum hardware would be an important step towards understanding the practical implications of isotropic errors.

\bmhead{Acknowledgements}
ASR acknowledges support of the German Federal Ministry of Research, Technology and Space (BMFTR)
within the framework programme ``Quantum technologies~\textendash~from basic research to market'' through the projects QSolid and QCStack. ASR also acknowledges support from the European Union through the Horizon Europe project OpenSuperQPlus, grant agreement number 101113946 and the European Innovation Council (EIC) Transition project QruiseOS, grant agreement number 101099538.

\bibliography{sn-bibliography}%

\end{document}